\documentclass{iaus}
\usepackage{graphicx}

\title[Stellar Parameters of Giants] 
{The Determination of Stellar Parameters of Giants in the Galactic
Disks and Bulge}

\author[Bystr\"om  et al.]   
{Joakim Bystr\"om$^1$, Nils Ryde$^1$, Sofia Feltzing$^2$, Johan
Holmberg$^3$ and Thomas Bensby$^4$}

\affiliation{$^1$Department of Astronomy and Space Physics, Uppsala
University, Sweden \break email: joabys, ryde@astro.uu.se\\[\affilskip]
$^2$Lund Observatory, Lund Universtiy, Sweden, email:
sofia@astro.lu.se\\[\affilskip]
$^3$Max Planck Institute for Astronomy, Germany, email:
holmberg@mpia.de\\[\affilskip]
$^4$Department of Astronomy University of Michigan, USA, email:
tbensby@umich.edu}

\pubyear{2007}
\volume{xxx}  
\pagerange{119--126}
\date{?? and in revised form ??}
\jname{Proceedings Title IAU Symposium}
\editors{A.C. Editor, B.D. Editor \& C.E. Editor, eds.}
\begin{document}

\maketitle

\begin{abstract}
Here, we present our on-going work on the determination of stellar
parameters of giants in the Galactic Disks and Bulge observed with
UVES on the VLT. We present some preliminarily results.

\end{abstract}


Our project aims at discerning whether the Galactic bulge and the
thick disk are evolutionarily connected, by differentially studying
their respective chemical evolution. Here, we present our initial
step dealing with the detailed determination of the fundamental
parameters of our target stars: metal-rich
($-0.5<\mathrm{[Fe/H]}<-0.1$), late-type ($4000\, \mathrm K <
\mathrm{T} < 5000\, \mathrm K$) giants. This step is important for
the subsequent step of accurately determining the abundances from
these rich spectra.

One advantage of our project is that it is a strictly differential
analysis of spectra, with homogeneously determined stellar
parameters. Thus, the stars will be analysed in an identical way in
order to firmly establish the picture of the abundance evolution of
the two populations in relation to each other. We will analyse
standard stars, thick disk and bulge stars.\\

Here we present the analysis of a sub-sample of our stars: 8
thick-disk and 8 standard stars. We are currently working on more
thick disk and  bulge giants. The spectra were recorded with UVES at
S/N$=75$ and $R=60\,000$, and were reduced with REDUCE (Piskunov \&
Valenti, 2002). The Fe line measurements were done in IRAF and the
line synthesis and analysis was done in the MARCS model-atmosphere
environment. Our analysis is based on the discussions in Fulbright
et al. (2006), who analysed bulge giants.

Equivalent widths for more than 120 Fe {\sc i} absorptions lines in
the middle and red parts of the spectra were measured. To be able to
determine the surface gravities also eight Fe {\sc ii} lines were
analysed.  By requiring that the iron abundance should be
independent of the excitation potential, we modified the temperature
of the models (increments of 25 K). The $\log g$ was estimated by
studying the difference in the abundance for Fe {\sc i} and Fe {\sc
ii}. Finally, diagrams with abundance vs $\log W/\lambda$ for Fe
{\sc i} were constructed to determine the microturbulence velocity.
When these parameters are defined it is possible to assign to the
star its metallicity. For the standard stars, it is possible to get
accurate starting values from photometry for $T_{\mathrm{eff}}$ and
$\log g$ and then iterate and find the spectroscopically based
parameters. For the thick disk stars more iterations were
required.\\


The photometrically and spectroscopically determined parameters for
the standard stars agree well, which is reassuring for our analysis
of the thick disk sample. Finally, we compared the spectroscopically
derived iron abundances for the thick disk stars with iron
abundances derived from DDO photometry using the calibration by
Holmberg \& Flynn (2004). The agreement is in general good. Table 1
summarizes our results.

Our full project includes three sets of stars, local standards,
thick disk stars, and stars in the Galactic bulge. The aim with the
current study is to investigate how well we can trust different
methods for derivation of the stellar parameters. For the Bulge
stars we will have significantly less good information as compared
to our data for the local standard stars.  Differential reddening
will be one of the greatest problems there. With the current results
it does appear that $\log g$ can be derived using excitation balance
in iron (see Table 1 for the standard stars). This is in agreement
with what was found by Kraft \& Ivans (2003).

We will also further investigate possible departures from the
assumption of LTE  in the derivation of the iron abundances in such
cool giant stars. For this we plan to use e.g. the methods and model
by Collet et al. (2005)

\begin{table}\def~{\hphantom{0}}
  \begin{center}
  \caption{Preliminary fundamental parameters of a sub-sample of our programme stars}
  \label{tab:kd}
  \begin{tabular}{lllllllrr}\hline
Star & $T_{\mathrm{eff}}$ & \# of & $T_{\mathrm{eff}}$ & $\log g$ &
$\log g$ & $\xi_{\mathrm{micro}}$ &[Fe/H] &[Fe/H]\\
& $\mathrm{(phot)}$ & colors & (spec) & (Hipp) &
($\mathrm{FeII/FeI}$) & [kms$^{-1}$] & (spec) & $\mathrm{(phot)}$\\
\hline
Standard stars:\\
HD113226 & 4974 (22) & 3 & 4961 &  2.80 & 2.80 &  1.5 &   0.07 \\
HD123139 & 4753 (51) & 1 & 4740 &  2.76 & 2.50 &  1.5 &  -0.15 \\
HD124897 & 4240 (37) & 7 & 4270 &  1.93 & 1.35 &  1.7 &  -0.62 \\
HD138716 & 4743 (43) & 4 & 4683 &  3.02 & 3.02 &  1.1 &  -0.14 \\
HD139663 & 4233 (15) & 2 & 4377 &  1.83 & 1.83 &  1.7 &   0.03 \\
HD161096 & 4503 (15) & 2 & 4534 &  2.48 & 2.46 &  1.5 &   0.16 \\
HD171443 & 4223 (51) & 1 & 4276 &  1.85 & 1.55 &  1.5 &  -0.12 \\
HD175190 & 4176 (21) & 6 & 4259 &  1.00 & 1.50 &  1.6 &  -0.22 \\
Thick Disk stars:\\
HD 1378  &4007 (51) &1 &4148 & & 1.00 & 1.5&  -0.10 & -0.13\\
HD 2763  &4169 (78) &3 &4246 & & 1.50 & 1.5&  -0.23 & -0.22\\
HD 3356  &4142 (16) &4 &4272 & & 1.25 & 1.5&  -0.50 & -0.35\\
HD 3524  &4016 (51) &1 &4029 & & 1.00 & 1.3&  -0.34 & -0.47\\
HD 3709  &4254 (9)  &4 &4327 & & 1.75 & 1.6&  -0.16 & -0.05\\
HD 4303  &4575 (22) &6 &4625 & & 1.75 & 1.5&  -0.20 & -0.14\\
HD 4955  &4043 (51) &1 &4135 & & 1.00 & 1.5&  -0.38 & -0.40\\
HD 8349  &4029 (51) &1 &4154 & & 1.25 & 1.4&  -0.20 & -0.27\\
\hline
  \end{tabular}
 \end{center}
\end{table}

\begin{acknowledgments}
We would like to thank Kjell Eriksson for useful discussions.
\end{acknowledgments}

\end{document}